\documentclass[conference]{IEEEtran}
\IEEEoverridecommandlockouts
\usepackage{cite}
\usepackage{amsmath,amssymb,amsfonts}
\usepackage{algorithmic}
\usepackage{graphicx}
\usepackage{textcomp}
\usepackage{xcolor}
\newcommand{\eg}{{\emph{e.g.}}}
\newcommand{\ie}{{\emph{i.e.}}}

\def\BibTeX{{\rm B\kern-.05em{\sc i\kern-.025em b}\kern-.08em
    T\kern-.1667em\lower.7ex\hbox{E}\kern-.125emX}}
\begin{document}

\title{Multi-scale alignment and Spatial ROI Module for COVID-19 Diagnosis\\}
\author{Hongyan Xu$^{1,2}$, Dadong Wang$^{2}$, Arcot Sowmya$^{1}$\\
$^1$School of Computer Science and Engineering, University of New South Wales, Kensington 2052, Australia\\
$^2$Data61, The Commonwealth Scientific and Industrial Research Organisation (CSIRO)\\
{\tt\small hongyan.xu@unsw.edu.au, Dadong.wang@csiro.au,a.sowmya@unsw.edu.au}
}

\maketitle

\begin{abstract}
  Coronavirus Disease 2019 (COVID-19) has spread globally and become a health crisis faced by humanity since first reported. Radiology imaging technologies such as computer tomography (CT) and chest X-ray imaging (CXR) are effective tools for diagnosing COVID-19. However, in CT and CXR images, the infected area occupies only a small part of the image. Some common deep learning methods that integrate large-scale receptive fields may cause the loss of image detail, resulting in the omission of the region of interest (ROI) in COVID-19 images and are therefore not suitable for further processing. To this end, we propose a deep spatial pyramid pooling (D-SPP) module to integrate contextual information over different resolutions, aiming to extract information under different scales of COVID-19 images effectively. Besides, we propose a COVID-19 infection detection (CID) module to draw attention to the lesion area and remove interference from irrelevant information. Extensive experiments on four CT and CXR datasets have shown that our method produces higher accuracy of detecting COVID-19 lesions in CT and CXR images. It can be used as a computer-aided diagnosis tool to help doctors effectively diagnose and screen for COVID-19.
\end{abstract}


\section{Introduction}

Coronavirus Disease 2019 (COVID-19) is an acute respiratory infectious disease that spread worldwide. According to a report by the Center for Systems Science and Engineering (CSSE) at Johns Hopkins University (JHU) \footnote{https://coronavirus.jhu.edu/map.html}, as of February 7, 2022, there were 394,674,835 confirmed cases worldwide, covering 214 countries and regions, with a staggering 5,738,604 deaths.


Early detection and rapid isolation are essential to effectively suppress the spread of the disease. Currently, real-time reverse transcription-polymerase chain reaction (RT-PCR) is the most commonly used method for clinical screening of COVID-19 patients \cite{yang2020imaging}. 
This method tests for COVID-19 infection by collecting respiratory samples such as throat and nasal swabs from suspected cases. Although RT-PCR can produce  results in a relatively short time, it can only produce a  binary result of positive or negative, and cannot obtain  the level of infection. On the other hand, for some cases with respiratory infection symptoms but negative RT-PCR test, radio imaging technologies such as computer tomography (CT) and chest X-ray imaging (CXR) tests for further diagnosis may be required \cite{salehi2020coronavirus}. Recent studies \cite{huang2020clinical} have found that most COVID-19 positive cases show similar features in CT and CXR images, \eg, ground-glass opacity and interstitial abnormalities. With these similar features, radiology imaging  can be a key step in diagnosing suspected patients. However, during early screening, it is challenging for radiologists to make a correct diagnosis in a short time due to similar manifestations of COVID-19 and other viral pneumonia. Therefore, an auxiliary algorithm that can accurately interpret radiology images and assist radiologists in diagnosing COVID-19 cases is essential.

\begin{figure}[t]\centering
\includegraphics[width=0.45\textwidth]{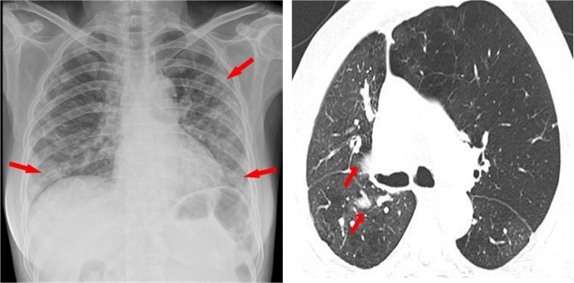}
\vspace{-2mm}
\caption{Examples of CXR (left) and CT (right) scans positive for COVID-19.  The red arrow in the picture indicates the lesions.} \label{fig1}
\vspace{-7mm}
\end{figure}

So far, some works on small-scale COVID-19 CT or CXR image datasets using deep learning technology for diagnosis have been reported \cite{ouyang2020dual}. While these works have shown promising accuracy, they suffer from two main limitations: (1) Most of the current works lack extensive comparison of model performance of on different datasets; (2) Most works do not propose a general algorithm that is suitable for different models and datasets, and are therefore limited in practical applications.

\begin{figure}[t]
\includegraphics[width=0.48\textwidth]{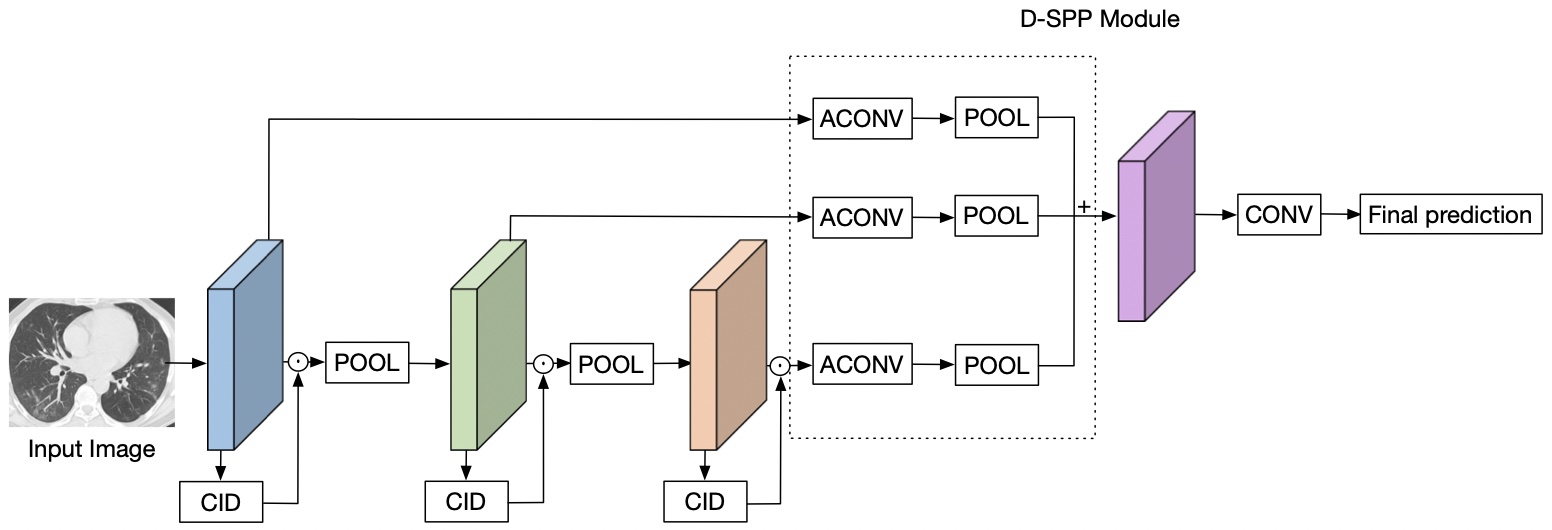}
\caption{Overall framework of proposed method.} \label{fig2}
\vspace{-5mm}
\end{figure}

To address the above problems, we propose a generic method that can be introduced into common image classification models \cite{su2021vision,xu2022data}(\eg, MobileNetV2, VGG19) to detect COVID-19 lesions in both CT and CXR images with improved performance. In detail, we introduce a deep spatial pyramid pooling (D-SPP) module to the front end of the network to integrate contextual information across regions and ensure the preservation of valid feature information in the images. To draw the network’s attention to areas related to disease-specific lesions, we also propose a COVID-19 infection detection (CID) module. The overall framework of the proposed method is shown in Fig. \ref{fig2}. Our main contributions are listed below.

\begin{enumerate}
    \item We propose a novel D-SPP module, which collects multi-scale image features and contextual information to guide subsequent accurate predictions. 
    \item We propose an CID module to analyse the region of interest (ROI) at  spatial level, thus drawing the neural network's attention to the ROI  while suppressing unrelated information in the images.
	\item We have validated the proposed method on four COVID-19 CT and CXR image datasets with many existing models, and compared the results with other COVID-19 detection algorithms \cite{alshazly2020explainable, jaiswal2020classification, panwar2020deep, yazdani2020covid, wang2020contrastive, ragab2020fusi, pathak2020deep, saqib2020covid19, ewen2020targeted, babukarthik2020prediction, ozdemir2020weighted}. 
	\item As plug-and-play modules, our D-SPP and CID modules can effectively boost the performance of common models for COVID-19 infection detection.
\end{enumerate}

The rest of the paper is organised as follows. Related work is reviewed in Section \ref{section2}, and Section \ref{section3} and \ref{section4} introduce the proposed method and describe the datasets used and the experiments conducted to evaluate the performance of the proposed method. In Section \ref{section5} the results of ablation studies are described, and Section \ref{section7} summarises the limitations of the study and possible future work, followed by conclusion in Section \ref{section6}.

\section{Related Work}\label{section2}
This section discusses several works most relevant to this work, including deep learning-based diagnosis of COVID-19, Atrous Spatial Pyramid Pooling (ASPP) module and attention mechanism.

\subsection{Deep Learning-Based Diagnosis of COVID-19}
In recent years, it has been shown that convolutional neural networks (CNN) can perform well in detecting lung diseases. In 2017, Rajpurkar \textit{et al.} proposed CheXNet \cite{rajpurkar2018deep}, which used a 121-layer CNN to detect 14 lung diseases in the ChestX-ray14 dataset \cite{wang2017chestx}. Gu \textit{et al.} \cite{gu2018classification} proposed a computer-aided diagnosis (CAD) system that can identify bacterial and viral pneumonia in chest radiography. 

Inspired by these achievements, recent studies have applied CNN to COVID-19 case diagnosis. Wang \textit{et al.} \cite{wang2020weakly} employed a weakly supervised deep learning framework for COVID-19 classification and lesion localisation, using 499 CT images for training and 131 CT images for testing,  and obtained an ROC (Receiver Operating Characteristic) AUC of 0.959, and a PR (Precision-Recall) AUC of 0.976. He \textit{et al.} \cite{he2020sample} introduced a Self-Trans method for COVID-19 diagnosis, verified on a COVID19-CT dataset containing 349 CT images, and achieved an F$_{1}$ score of 0.85 and an AUC of 0.94. Wang \textit{et al.} \cite{wang2020tailored} proposed  COVID-Net for COVID-19 detection on chest X-rays, which achieved an accuracy of 93.3\% on a dataset containing 13,975 CXR images. 

\subsection{Atrous Spatial Pyramid Pooling (ASPP) Module}
Modern image classification networks integrate multi-scale context information through continuous pooling and down-sampling layers, resulting in loss of detailed information about the object edges and degradation of the image resolution \cite{simonyan2014very}. To address this problem, He \textit{et al.} \cite{he2015spatial} proposed a layered Spatial Pyramid Pooling (SPP) module which can obtain the fusion of information and receptive fields from different sub-regions. Experiments show that multi-scale feature information fusion can bring about improvement of network accuracy. This is not simply because of the increase in parameters, but because multi-level pooling can effectively deal with object deformation and differences in the spatial layout.
Compared with the SPP module, the atrous convolution used by the Atrous Spatial Pyramid Pooling (ASPP) module avoids the loss of image detail information caused by the down-sampling operation; therefore, it fits better with the need to detect COVID-19 lesions with a small range in the image. The ASPP module is part of the DeepLabv2 \cite{chen2017deeplab} model. It performs parallel atrous sampling at different sample rates on a given input, equivalent to capturing the context of an image on multiple scales, resulting in multi-scale image feature information. 
\subsection{Squeeze-and-Excitation (SE) Module}
The Squeeze-and-Excitation (SE) module was proposed by Hu \textit{et al.} \cite{hu2018squeeze}, which is designed to explicitly model the interdependence between channels of its convolutional features, thereby improving model performance. The SE module consists of two parts. The first part is the squeeze module, which compresses feature maps along the spatial dimensions to generate feature descriptors and obtains the global distribution of channel-wise responses. The second part is the excitation module, which explicitly models the correlation between feature channels to generate channel-wise weights. With these two modules, the quality of the network's feature representation is improved, since the interdependencies between the channels of its convolutional features are explicitly modelled. In general, the SE module can achieve feature recalibration. Learning global information makes it possible to emphasise important features and suppress less important features selectively.

\subsection{Attention Mechanism}
An attention mechanism can be interpreted as a means of allocating available computing resources to the most informative components of a signal \cite{hu2018squeeze}. It can help the model assign different weights to each part of the input, extract more critical information and enable the model to make more accurate judgment, without incurring high costs for model computation and storage. Therefore, it is widely used in many fields such as machine translation \cite{bahdanau2014neural}, image captioning \cite{xu2015show}, and text summarisation \cite{rush2015neural}. Some studies have been carried out on spatial and channel-related attention mechanisms. By modelling the interdependencies between channels, SENet can improve the quality of the representations generated by the network and enable feature recalibration. The Convolutional Block Attention Module (CBAM) \cite{woo2018cbam} uses two modules, namely Channel Attention Module (CAM) and Spatial Attention Module (SAM), to sequentially infer attention maps along two separate dimensions (channel and space). In this work, we are committed to paying attention to the spatial dimensions and giving different attention levels to different areas of the feature map. This allows the network to focus more on the COVID-19 lesion areas, suppress the influence of unrelated areas and make effective decisions.

\section{Proposed Method}\label{section3}
In this section, we shall detail the components of the proposed approach for COVID-19 infection detection, including the deep spatial pyramid pooling (D-SPP) module and the COVID-19 infection detection (CID) module.

\subsection{Deep Spatial Pyramid Pooling (D-SPP) Module} \label{section_shizhu1}
Inspired by the ASPP module, we propose a deep pyramid pooling module named D-SPP. For CT or CXR images used for COVID-19 diagnosis, the affected area often only occupies a small part of the image, as shown in Fig. \ref{fig1}. 

In the early stage, the affected area appears as ground-glass opacity (GGO), and at a later stage, it appears as lung consolidation. Observing the affected area locally reveals a small range of detailed features, and for the entire image, the affected area shows regional characteristics. Both types of information are needed to diagnose COVID-19 accurately. 
However, for networks, over-down-sampling can result in loss of critical detail. Therefore, we add the D-SPP module to the front end of the network, extracting feature maps of different sizes across regions to predict the  overall image characteristics at the front end of the network and help the network better understand the image. The structure of the proposed D-SPP module is shown in Fig. \ref{fig2}.

\begin{figure}
\centering
\includegraphics[width=0.35\textwidth]{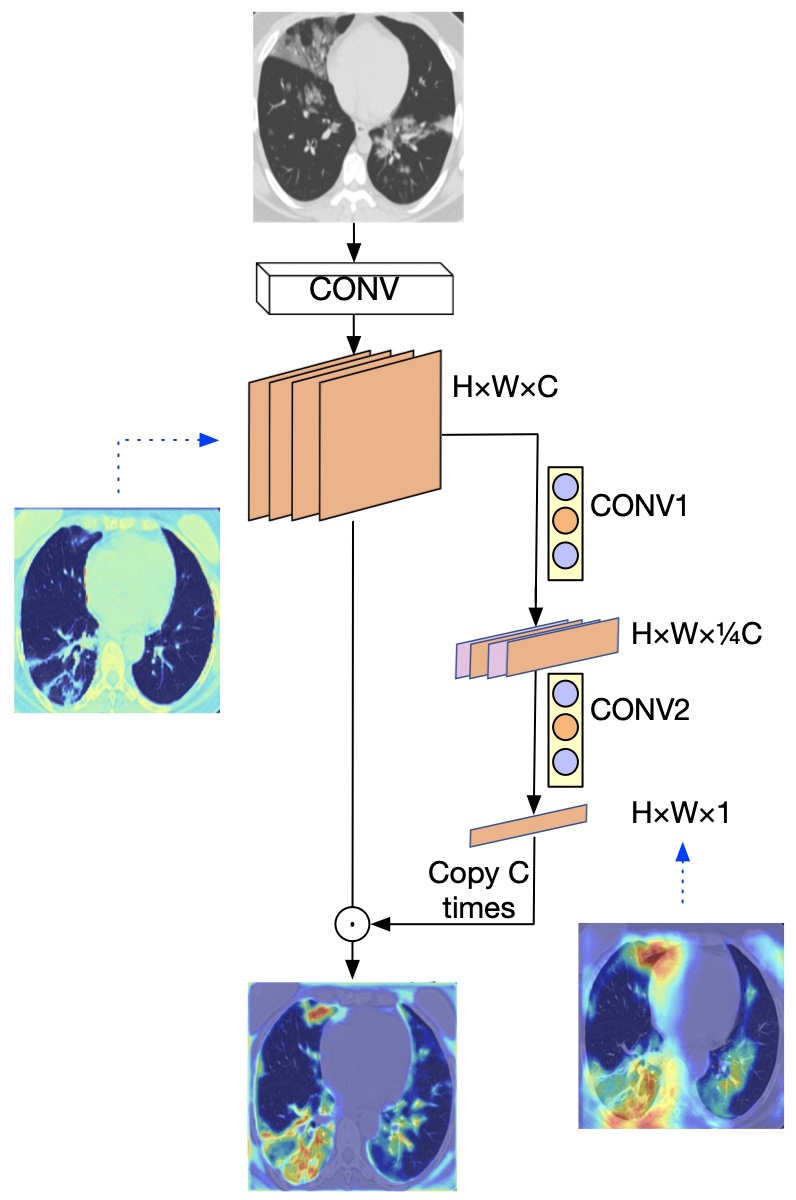}
\caption{CID Module.} \label{fig3}
\end{figure}

Unlike the original ASPP module, which uses the same layer feature map to integrate the context information of multiple received fields, the proposed D-SPP module takes the feature maps of different layers for sampling. Specifically, if a network is trained on the ImageNet dataset, it will undergo five down-sampling processes. In our method, we used the feature maps before down-sampling as the input of the D-SPP module, and applied different atrous rates to them. Then, to effectively extract information from feature maps with different resolutions, we adopt the atrous rate directly related to the size of the feature map. In other words, the misalignment issue caused by merging feature maps of  different resolutions is thereby solved. Suppose the size of the feature map is $[H_i, W_i, D_i]$ for the $i$-th input, and we apply atrous convolution with an atrous rate of $\mathcal{R}_i$. Then, we can model $\mathcal{R}_i$ with function $f(H_i,W_i)$, and we have
\begin{equation}
    \frac{\mathcal{R}_i}{\mathcal{R}_j} = \frac{f(H_i,W_i)}{f(H_j,W_j)}
    \label{resolution1}
\end{equation}
In practice, we usually specify a minimum atrous rate as $\alpha$, then the atrous rate for each part in D-SPP can be formulated as
\begin{equation}
    \mathcal{R}_i = \alpha \times \frac{f(H_i,W_i)}{f_{min}(H_{min},W_{min})}
    \label{resolution2}
\end{equation}
To simplify, we can directly use the linear function for $f$, so Eq. \eqref{resolution2} can be rewritten as
\begin{equation}
    \mathcal{R}_i = \alpha \times \sqrt{\frac{H_i~W_i}{H_{min}~W_{min}}}
    \label{resolution3}
\end{equation}

In fact, the minimum $\alpha$ determines the spatially fused information of the receptive field in feature maps. In contrast, the second term in Eq. \eqref{resolution3} ensures that all the feature maps across different resolutions have the same receptive field.

\subsection{COVID-19 infection detection(CID) Module}


In deep neural networks, the size of the receptive field can roughly indicate the extent to which context information is used \cite{zhao2017pyramid}. However, Zhou \textit{et al.} \cite{zhou2014object} showed that the empirical receptive field of CNN is much smaller than the theoretical receptive field, especially in the higher layers. The SE module \cite{hu2018squeeze} is an effective way to address this  problem. It is a feature-level attention mechanism that aggregates all pixels of the feature map through average pooling to obtain a global receptive field. The information of different feature maps is then extracted and fused using a $1 \times 1$ convolution. However, this simple average pooling will treat all pixel information in the image equally. Unlike other classification tasks \cite{xu2019automatic,zoph2018learning, zagoruyko2016wide,su2021bcnet,su2021k,su2021locally,su2021prioritized}, the features related to COVID-19 infection in the COVID-19 detection task are usually restricted to a small area. In other words, most of the image areas are irrelevant to the task, which may negatively affect accuracy. Therefore, by directly integrating global information through average pooling in the SE module, the effective information will be submerged in a large amount of invalid information.

Therefore, the key to improving the accuracy of COVID-19 detection is to make the network pay more attention to the areas associated with COVID-19 infection, and ignore unrelated areas as much as possible. We introduce a new COVID-19 task-oriented module named the CID module. Unlike the channel-wise attention module proposed in SENet, the CID module aims to highlight the important areas within feature maps in the spatial dimension. We propose to construct pixel-level attention for all feature maps. To highlight valid information and suppress irrelevant areas, we need to simultaneously re-scale the pixel weights for all feature maps. In detail, for the input feature map with size $[H_{f},W_{f},D_{f_{i}}]$, after passing through the CID module, the size of the feature map becomes $[H_{f},W_{f},D_{f_{j}}]$. We define the attention of the feature map to pixels as the Attention factor $A_f$, which can be calculated as follows:
\begin{equation}\label{eq1}
   A_f =  \frac{D_{f_{i}}}{D_{f_{j}}}
\end{equation}
 In particular, when $D_{f_{j}}=1$, we pay the same attention to pixels at the same position in all the feature maps.
	\begin{figure}
	\centering
		\includegraphics[width=0.38\textwidth]{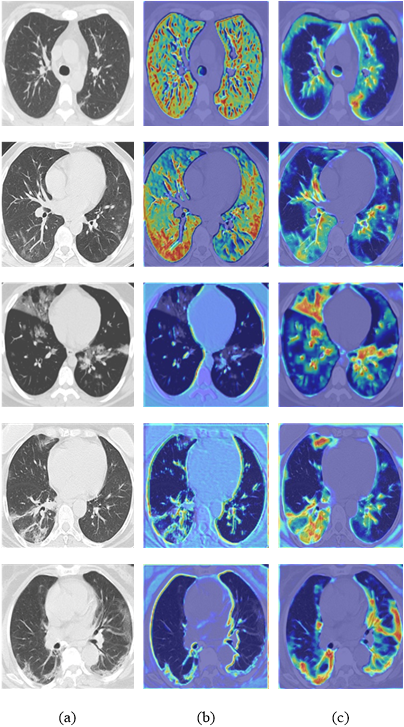}
		\caption{Heat map of COVID-19 CT images.\label{fig4}}
\end{figure} 
 
As shown in Fig. \ref{fig3}, suppose the original input CT or X-ray image is of size $[H, W, C]$. Then in the CID module, through the convolution kernel of size $1 \times 1 $ and an output channel of size c/4, a feature map of size $[H, W, C/4]$ is generated. The output feature map is further processed by a convolution kernel of size  $1 \times 1 $ and an output channel of size 1 to generate an $[H, W, 1]$ size feature map, which is used as a spatial level attention feature map to analyse the important areas in all feature maps. Then, we highlight the original input by copying it $C$ times, and directly dot multiplying them with each other. As a result, using the CID module, the network can pay more attention to highlighted areas, \ie, COVID-19 infected regions of CT and X-ray images, while suppressing irrelevant information such as bony areas or image boundaries. 

 \subsection{Visual Analysis} \label{section_sxshizhu2}

To better demonstrate the effectiveness of the proposed method in focussing on the COVID-19 infected area in CT images, we visualised the learned attention region generated by the ResNet50 model with or without the proposed D-SPP and CID modules respectively, as shown in Fig. \ref{fig4}. For each CT volume, Grad-CAM \cite{selvaraju2017grad} visualisation was applied for visual comparison. Column (a) corresponds to CT images of patients infected with COVID-19, and Column (b) corresponds to the results generated by the ResNet50 model without the D-SPP and CID modules, referred to below as the general ResNet50 model. Column (c) is generated by the ResNet50 model with the D-SPP and CID modules. In Column (a), the COVID-19 effect appears as an abnormal shadow area in the CT image. By comparing Columns (b) and (c), we note that the general ResNet50 model either does not locate the area associated with COVID-19 infection or misrecognises almost the entire lung as a potential infection area associated with COVID-19, which is not conducive to the detection of COVID-19 infected areas. On the contrary, the ResNet50 model with D-SPP and CID modules can accurately identify the locations of potential lesions and assign a larger coefficient to the relevant areas of the feature map, allowing the network to focus on areas affected by COVID-19 (\eg, ground glass-like areas in the image) rather than unrelated areas. In other words, the addition of the D-SPP and CID modules makes disease localisation more accurate. Experimental results of the proposed approach on the CT dataset are presented in Table \ref{table2}.

\begin{figure}
\centering
		\includegraphics[width=0.38\textwidth]{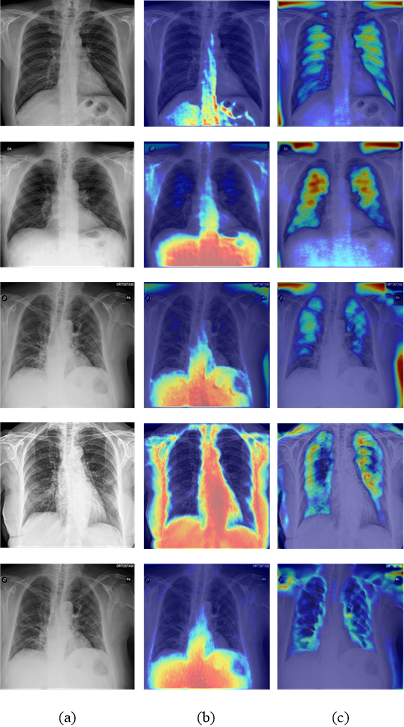}
		\caption{Heat map of COVID-19 X-Ray images.\label{fig5}}
\end{figure}

To demonstrate the capability of the proposed module for the detection of COVID-19 infected areas in chest X-ray images, we visualised the learned attention regions generated by the ResNet50 model with or without the proposed D-SPP and CID modules on the X-ray dataset, as shown in Fig. \ref{fig5}. Similar to Fig. \ref{fig4}, Column (a) in Fig. \ref{fig5} shows the chest X-ray images of patients infected with COVID-19, Column (b) corresponds to the test results of the general ResNet50 model, Column (c) corresponds to the results generated by ResNet50 model with the D-SPP and CID modules. In Column (a), the COVID-19 affected areas are represented as a shadowed area of the lungs. By comparing Column (b) and (c), we can find that the general ResNet50 model focusses on areas of the X-ray images unrelated to COVID-19, such as the spine region. After adding the proposed modules, the model is focussed on the COVID-19 affected lung areas, or more specifically the shadow areas of the lungs. This shows that the D-SPP and CID modules can maintain focus on the COVID-19 affected areas and make evidence-based decisions. The relevant experimental results are presented in Table \ref{table3}.

\section{Experiments and Analysis}\label{section4}
To verify the effectiveness of the proposed method, we conducted extensive experiments on three COVID-19 CT and X-ray image datasets, as shown in Table \ref{tabel111}. For all experiments, we set $\alpha$ to 3, and detailed experimental settings are elaborated in Section \ref{experimental_settings}.
\subsection{Benchmark Datasets}

\subsubsection{SARS-CoV-2 CT Scan Dataset \cite{soares2020sars}}
This dataset consists of 2,482 CT images, including 1,252 COVID-19-positive images and 1,230 images of pulmonary diseases that are not COVID-19.  The data were collected from a hospital in Sao Paulo, Brazil. 

\subsubsection{COVID19-CT Dataset \cite{he2020sample}} This dataset contains 349 positive CT scans with clinical manifestations of COVID-19 and 397 negative CT scans with no COVID-19 detected. It was constructed by Yang \textit{et al.} by collecting medical images in COVID-19-related medRxiv and bioRxiv papers. 

\subsubsection{COVID X-ray Dataset}
There are two sources for this dataset. One is the IEEE dataset available on Github
\footnote{https://github.com/ieee8023/covid-chestxray-dataset}, and the author only intercepts data related to COVID-19; the other is normal data from the Kaggle dataset
\footnote{https://www.kaggle.com/paultimothymooney/chest-xray-pneumonia}. The original dataset contains 98 images, including 78 X-rays of COVID-19-positive patients and 20 X-rays of normal persons. For the experiments, we divided it into training and test sets in the ratio 78:20. 

\begin{table}[t]\centering
\caption{Four COVID-19 CT and X-ray image datasets we used.}\label{tabel111}
\begin{tabular}{c|c|c|c}
\hline
Datasets      & Type  & COVID Images & Normal Images \\ \hline
SARS-CoV-2 CT & CT    & 1252         & 1230          \\
COVID19-CT    & CT    & 349          & 397           \\
COVID X-Ray   & X-ray & 78           & 20            \\
COVID-Xray-5k & X-ray & 71           & 5000          \\ \hline
\end{tabular}
\end{table}

\subsubsection{COVID-Xray-5k Dataset \cite{minaee2020deep}} This dataset consists of 2,031 training images and 3,040 test images. Among them are 71 COVID-19-positive X-rays and 5,000 COVID-19-negative X-rays. The X-ray images in this   dataset come from two datasets, namely Covid Chestxray-Dataset \footnote{https://github.com/ieee8023/covid-chestxray-dataset} and ChexPert dataset \cite{irvin2019chexpert}.

\subsection{Models and Experimental Settings} \label{experimental_settings}
In this section, we provide details of the experiments conducted. In general, for all datasets and different models, we use SGD optimizer with momentum 0.9. We set weight decay to $10^{-4}$ for ResNet50 and VGG19, and to $5 \times 10^{-5}$ for MobileNetV2. The learning rate is decayed with cosine from 0.1 to $10^{-5}$ for all models. To accommodate the differing sizes different datasets, we set batch size to 32 for the SARS-CoV-2 CT scan dataset and COVID-Xray-5k dataset, 16 for the COVID19-CT and 8 for COVID X-ray dataset. Besides, all models are trained for 100 epochs on all datasets. 

All experiments were performed on an Intel(R) Core(TM) i9-7980XE CPU@ 2.60GHz CPU (Santa Clara, USA, Intel) workstation with a 16GB RAM and NVIDIA 2080Ti$\times2$ GPU. The CNN  was constructed using Pytorch. 

To quantitatively evaluate the performance of the proposed model, we utilise several evaluation metrics commonly used in classification tasks: (1) Accuracy: ratio of the number of instances classified correctly to the total number of instances,  representing the overall effectiveness of the classifier; (2) Precision: out of all the predicted positive instances, the percentage that represents true positive instances; (3) Recall (also called Sensitivity): out of all the positive instances, the percentage that is identified correctly, representing the effectiveness of the classifier in identifying positive instances; (4) F$_{1}$ score: the harmonic mean of Precision and Recall, measuring the relationship between the positive label of the data and the label given by the classifier; (5) AUC (Area Under the Curve): area under the receiver operating characteristic (ROC) curve, which indicates the false positive rate change against the true positive rate change.

\subsection{Experiments on SARS-CoV-2 CT Scan Dataset}
 Consistent with other work\cite{soares2020sars}, we divided the dataset into training and validation sets at a scale of 80\%:20\%. We added the proposed D-SPP module and CID module to currently popular clasification models, namely MobileNetV2, ResNet18, ResNet34, ResNet50 and VGG19, for experiments. 
 
 To verify the effectiveness of the proposed method, we compare it with other models, including the explainable Deep Learning approach (xDNN) \cite{soares2020sars}, the DenseNet201 model \cite{jaiswal2020classification}, and so on. As shown in Table \ref{table2},  MobileNetV2, ResNet18, ResNet34, ResNet50, and VGG19 models with the proposed D-SPP and CID modules surpass the xDNN model in accuracy, recall and F$_{1}$ score. Compared with other models in Table \ref{table2}, the ResNet50 model using our method is slightly lower than the ResNet101 model in terms of precision. Nevertheless, it performs best overall in terms of accuracy, recall and F$_{1}$ score, reaching over 99\%.
 
 \begin{table*}[]
 \vspace{-6mm}
 \caption{Performance Comparison on CT datasets.}\label{table2}
 \centering
 \setlength{\tabcolsep}{4mm}{
\begin{tabular}{c|cccccc}
\hline
Dataset                    & Model               & Accuracy         & Precision        & Recall           & F$_{1}$ Score         & AUC              \\ \hline
                           & xDNN\cite{soares2020sars}                & 97.38\%          & 99.16\%          & 95.53\%          & 97.31\%          & 97.36\%          \\
                           & MobileNetV2\_ours   & 99.20\%          & 99.22\%          & 99.21\%          & 99.21\%          & 98.05\%          \\
                           & ResNet18\_ours      & 99.40\%          & 98.79\%          & 99.61\%          & 99.20\%          & 98.32\%          \\
                           & ResNet34\_ours      & 99.40\%          & 99.20\%          & 99.61\%          & 99.40\%          & 98.74\%          \\
                           & ResNet50\_ours      & \textbf{99.61\%} & 99.21\%          & \textbf{100\%}   & \textbf{99.60\%} & 99.14\%          \\
                           & VGG19\_ours         & 98.60\%          & 98.02\%          & 99.21\%          & 98.61\%          & 97.22\%          \\
                           & ResNet50 \cite{alshazly2020explainable}            & 99.20\%          & 99.10\%          & 99.40\%          & 99.20\%          & /                \\
SARS-COV-2 CT 
& ResNet101 \cite{alshazly2020explainable}           & 99.40\%          & \textbf{99.60\%} & 99.10\%          & 99.40\%          & /                \\
                           & DenseNet201 \cite{jaiswal2020classification}         & 96.25\%          & 96.29\%          & 96.29\%          & 96.29\%          & 97.00\%          \\
                           & Modified VGG19 \cite{panwar2020deep}      & 95.00\%          & 95.30\%          & 94.00\%          & 94.30\%          & /                \\
                           & COVID CT-Net \cite{yazdani2020covid}       & /                & /                & 85.00\%          & 96.20\%          & 97.00\%          \\
                           & Contrasive Learning \cite{wang2020contrastive} & 90.83\%          & 95.75\%          & 85.89\%          & 90.87\%          & 96.24\%          \\
                           & ShuffleNet \cite{ragab2020fusi}          & 96.30\%          & 96.00\%          & 97.00\%          & 96.50\%          & 99.00\%          \\
                           & ResNet18 \cite{ragab2020fusi}            & 97.60\%          & 97.50\%          & 97.50\%          & 97.50\%          & \textbf{100\%}   \\
                           & DL FUSION \cite{ragab2020fusi}           & 98.60\%          & 99.00\%          & 98.10\%          & 98.60\%          & \textbf{100\%}   \\
                           & CNN \cite{pathak2020deep}                 & 98.37\%          & 98.74\%          & 98.87\%          & 98.14\%          & 98.32\%          \\ \hline
                           & ResNet50 \cite{he2020sample}            & 69.00\%          & /                & /                & 72.00\%          & 76.00\%          \\
                           & MobileNetV2\_ours   & 82.76\%          & 82.51\%          & 81.63\%          & 82.07\%          & 84.33\%          \\
                           & ResNet50\_ours      & \textbf{85.22\%} & \textbf{84.73\%} & 84.69\%          & 84.71\%          & \textbf{88.12\%} \\
                           & VGG19\_ours         & 80.79\%          & 80.38\%          & 79.59\%          & 79.98\%          & 82.51\%          \\
COVID19-CT  & VGG19 \cite{saqib2020covid19}               & 80.30\%          & 78.76\%          & \textbf{84.76\%} & 81.65\%          & 87.96\%          \\
                           & ResNet50 \cite{saqib2020covid19}            & 80.79\%          & 83.00\%          & 79.05\%          & 80.98\%          & 87.69\%          \\
                           & MobileNetV2 \cite{saqib2020covid19}         & 76.85\%          & 77.36\%          & 78.10\%          & 77.73\%          & 85.49\%          \\
                           & Contrasive Learning \cite{wang2020contrastive} & 78.69\%          & 78.02\%          & 79.71\%          & 78.83\%          & 85.32\%          \\
                           & InceptionV3 \cite{ewen2020targeted}         & 84.23\%          & /                & /                & 85.05\%          & 84.15\%          \\
                           & DenseNet169 \cite{ewen2020targeted}         & 84.24\%          & /                & /                & \textbf{85.32\%} & 84.08\%          \\ \hline
\end{tabular}}
\end{table*}

\begin{table*}[]
\vspace{-4mm}
\caption{Performance Comparison on X-ray datasets.}\label{table3}
\centering
\setlength{\tabcolsep}{5mm}{
\begin{tabular}{c|cccccc}
\hline
Dataset                                                       & Model             & Accuracy         & Precision      & Recall         & F$_{1}$ Score         & AUC            \\ \hline
                                                              & VGG16             & 100\%            & 92.86\%        & 100\%          & 96.00\%          & /              \\
\begin{tabular}[c]{@{}c@{}}COVID X-ray\end{tabular} & MobileNetV2\_ours & 100\%            & \textbf{100\%} & 100\%          & \textbf{100\%}   & \textbf{100\%} \\
                                                              & ResNet50\_ours    & 100\%            & \textbf{100\%} & 100\%          & \textbf{100\%}   & \textbf{100\%} \\
                                                              & VGG19\_ours       & 100\%            & \textbf{100\%} & 100\%          & \textbf{100\%}   & \textbf{100\%} \\ \hline
                                                              & ResNet18\cite{minaee2020deep}          & /                & 97.50\%        & 88.80\%        & 92.90\%          & /              \\
                                                              & ResNet50\cite{minaee2020deep}          & /                & 97.50\%        & 90.50\%        & 93.90\%          & /              \\
                                                              & SqueezeNet\cite{minaee2020deep}        & /                & 97.50\%        & 97.80\%        & 97.60\%          & /              \\
                                                              & DenseNet121\cite{minaee2020deep}       & /                & 97.50\%        & 81.30\%        & 88.70\%          & /              \\
                                                              & MobileNetV2\_ours & 99.84\%          & 97.50\%        & 99.73\%        & 98.60\%          & 98.72\%        \\
\begin{tabular}[c]{@{}c@{}}COVID\_5k\end{tabular}   & ResNet50\_ours    & \textbf{99.93\%} & 97.50\%        & 99.90\%        & \textbf{98.69\%} & \textbf{100\%} \\
                                                              & VGG19\_ours       & 99.77\%          & 95.00\%        & 99.67\%        & 97.28\%          & 97.74\%        \\
                                                              & GDCNN \cite{babukarthik2020prediction}             & 98.84\%          & 93.00\%        & \textbf{100\%} & 96.37\%          & \textbf{/}     \\
                                                              & ACNN \cite{ozdemir2020weighted}              & 87.42\%          & \textbf{/}     & 75.00\%        & \textbf{/}       & \textbf{/}     \\
                                                              & ResNet50 \cite{ozdemir2020weighted}          & 94.97\%          & \textbf{/}     & 90.00\%        & \textbf{/}       & \textbf{/}     \\ \hline
\end{tabular}}
\end{table*}

\begin{figure}

		\includegraphics[width=0.45\textwidth]{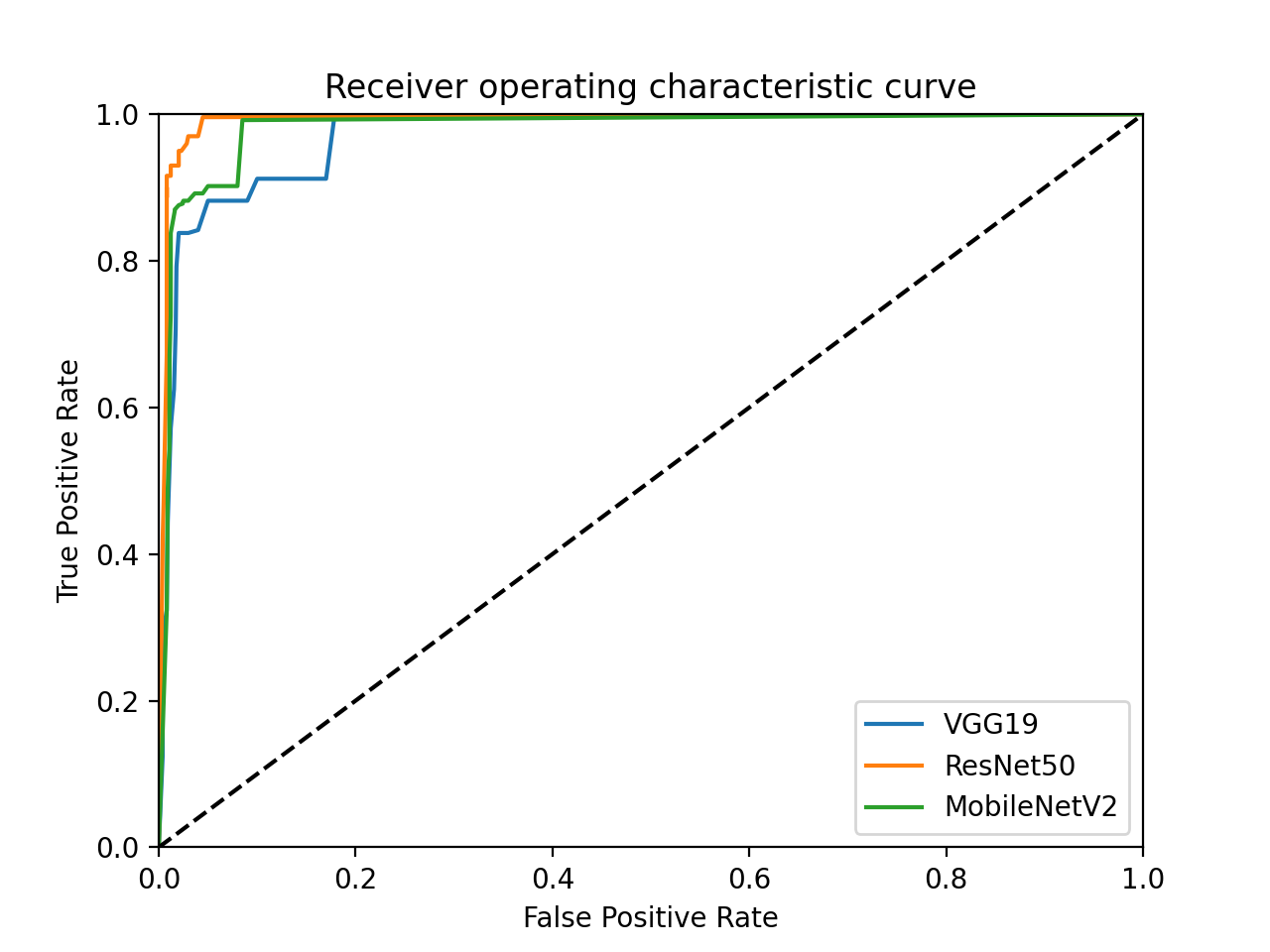}
		\centering
		\caption{ROC curves of three models on SARS-CoV-2 CT scan dataset.\label{fig6}}
\end{figure}

The ROC curves of VGG19, ResNet50 and MobileNetV2 models using our method on the SARS-CoV-2 CT scan dataset are shown in Fig. \ref{fig6}. The ResNet50 model has the best performance. The AUC values of the three models are 97.22\%, 99.14\% and 98.05\% respectively. The VGG19 model has a true positive rate (TPR) of 0.992 when the false positive rate (FPR) value is 0.178; the MobileNetV2 model has a TPR of 0.992 when the FPR value is 0.085; The ResNet50 model obtains the best performance with a TPR value of 0.996, with FPR value of about 0.045, among the three models. TPR approaching one means that the model can accurately identify COVID-19 positive samples, while a low FPR means that only a few negative samples are misjudged as positive samples. This fully demonstrates the effectiveness of the proposed method for the detection of COVID-19 from the SARS-CoV-2 CT scan images.




\begin{figure}
		\includegraphics[width=0.45\textwidth]{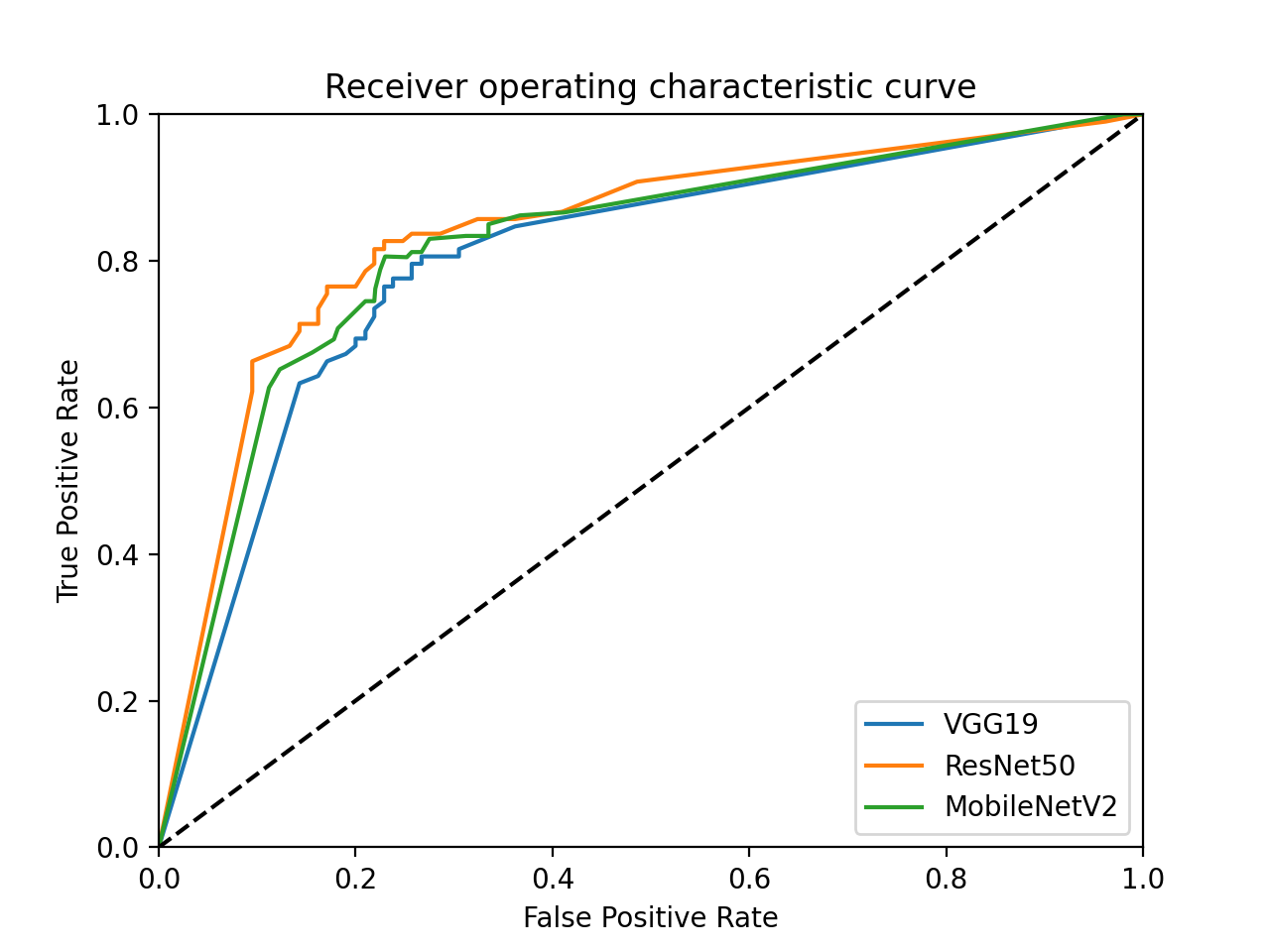}
		\centering
		\caption{ROC curves of three models on COVID19-CT dataset.\label{fig7}}
\end{figure}

 \subsection{Experiments on COVID19-CT Dataset}
 Consistent with other work \cite{he2020sample}, the COVID19-CT dataset was divided into a training, validation set and  test set in the ratio 0.6:0.15:0.25. We conducted experiments using ResNet50, MobileNetV2 and VGG19 models with our D-SPP and CID modules. Then we compared them with the ResNet50 model used elsehwere \cite{he2020sample} and the VGG19 model used in related work \cite{saqib2020covid19}, and the results are shown in Table \ref{table2}. The three models using our method are superior to the ResNet50 model results \cite{he2020sample}. In addition, the accuracy, F$_{1}$ score  and AUC achieved by the Resnet50 model integrated with the proposed method are respectively 16.22\%, 12.71\%, and 12.12\% higher than the ResNet50 used alone \cite{he2020sample}. In general, although recall and AUC are slightly lower than the VGG19 model results \cite{saqib2020covid19} and the DenseNet169 model \cite{ewen2020targeted}, our enhanced ResNet50 model outperforms the other models in Table \ref{table2} in terms of accuracy, precision and F$_{1}$ score.

The performance of VGG19, ResNet50 and MobileNetV2 models enhanced with our method on the COVID19-CT dataset is shown in Fig. \ref{fig7}. The AUC values of these three models are 82.51\%, 88.12\% and 84.33\% respectively. All three models have TPR greater than 0.8 when the FPR value is 0.4. The Resnet50 model has the best performance among the three models. 

\subsection{Experiments on COVID X-ray Dataset}
To assess the performance of our amethod on COVID-19 X-ray images, we conducted experiments using ResNet50, MobileNetV2 and VGG19 models that are integrated with our proposed D-SPP and CID modules on the COVID X-ray dataset. A comparison between the experimental results of our model and the VGG16 model used by the authors of the dataset is shown in Table \ref{table3}. The accuracy, precision, recall and F$_{1}$ score of our model are all 100\%. On the one hand, this is because the dataset is relatively small, as it only contains 98 images. On the other hand, it also shows the effectiveness of the proposed D-SPP and CID modules.

\subsection{Experiments on COVID-Xray-5k Dataset}
Separating COVID-19 positive X-ray images from other lung diseases and normal X-ray images is essential for COVID-19 diagnosis. So we conducted experiments on the COVID-Xray-5k dataset, which contains X-ray images of normal lung and 13 lung diseases. As shown in Table \ref{table3}, compared with the models used elsewhere \cite{minaee2020deep}, our model achieves great improvement in recall and F$_{1}$ score. The recall and F$_{1}$ score achieved by the ResNet50 model with our method were 9.40\% and 4.79\% higher than the original Resnet50 model respectively. Compared to the GDCNN model \cite{babukarthik2020prediction}, our Resnet50 model is 0.10\% lower in recall, but 1.09\%, 4.50\% and 2.32\% higher in accuracy, precision, and F$_{1}$ score respectively. The experimental results suggest that the Resnet50 model enhanced with D-SPP and CID modules is superior to the other models in Table \ref{table3} on accuracy, F$_{1}$ score and AUC. 

\section{Ablation Studies}\label{section5}
\textbf{Effect of the position of D-SPP and CID modules}   \quad To test the impact of the location of the D-SPP module and the presence of the CID module on the accuracy of the model, we conducted further experiments, the results of which are shown in Table \ref{table4}. Stage i (i = 4, 5, 6) indicates that the D-SPP is placed after the i-th stage of the model, i.e. the output of the i-th stage is used as the input of the D-SPP module. MobileNetV2 and ResNet18 were used as the backbone networks in these experiments.

\begin{table}[]\centering
\vspace{-6mm}
\caption{Ablation Experiments on SARS-CoV-2 CT scan dataset.}
\setlength{\tabcolsep}{0.3mm}{
\begin{tabular}{c|c|c|cc|cc}
\hline
\multicolumn{3}{c|}{Location of D-SPP} & \multicolumn{2}{c|}{CID}      & \multicolumn{2}{c}{Accuracy of models} \\ \hline
Stage 4       & Stage 5       & Stage 6       & \multicolumn{1}{c|}{with} & without & MobileNetV2        & ResNet18           \\ \hline
              &               &               & \multicolumn{1}{c|}{}     & \checkmark       & 96.87\%            & 97.57\%            \\
              &               &               & \multicolumn{1}{c|}{\checkmark}    &         & 97.16\%            & 97.89\%            \\
              &               & \checkmark             & \multicolumn{1}{c|}{}     & \checkmark       & 97.52\%            & 98.02\%            \\
              &               & \checkmark             & \multicolumn{1}{c|}{\checkmark}    &         & 98.19\%            & 98.39\%            \\
              & \checkmark             & \checkmark             & \multicolumn{1}{c|}{\checkmark}    &         & 98.59\%            & 98.79\%            \\
\checkmark             & \checkmark             & \checkmark             & \multicolumn{1}{c|}{\checkmark}    &         & \textbf{99.20\%}   & \textbf{99.40\%}   \\ \hline
\end{tabular}}
\vspace{-6mm}
\label{table4}
\end{table}

It can be observed that the addition of D-SPP and CID modules does improve the accuracy of the MobileNetV2 and ResNet18 models. Even if we only use the CID module, the accuracy of the MobileNetV2 model and ResNet18 model are slightly higher than the original model. When we add the D-SPP module to Stage 6 and do not use CID module, the accuracy of the two models is slightly improved compared with only using the CID module. When we add the D-SPP module in Stages 4, 5 and 6, and use the CID module, the enhanced MobileNetV2 and ResNet18 models have the best accuracy, which is 2.33\% and 1.83\% higher than the original models respectively. Experimental evidence suggests that addition of the D-SPP and CID modules does improve model performance.

\textbf{Effects of pre-training} \quad Due to the restriction on the sharing of COVID-19 images, many COVID-19 image datasets contain only a small number of images. Training and testing on a small dataset alone will limit the model generalisation. We, therefore, wanted to explore the impact of pre-training on an additional COVID-19 dataset on model performance on the current dataset. After pre-training on the SARS-CoV-2 CT dataset, we fine-tuned the model on the COVID19-CT dataset. In Table \ref{table5} the performance of the ResNet50 backbone without pre-training, and with pre-training and fine-tuning under different settings are shown. In the table, 40$+$pretrain40 means pre-training for 40 epochs on the SARS-CoV-2 CT dataset, followed by fine-tuning for 40 epochs on the COVID19-CT dataset; the rest can be deduced by analogy. We make the following observations: (1) Pre-training on the SARS-CoV-2 CT dataset can significantly improve model performance on the COVID19-CT dataset. Compared with the original model, even the ResNet50 backbone with 40$+$pretrain40 achieved an improvement of 2.46\%, 0.85\%, 3.06\% and 1.96\% on the accuracy, precision, recall and F$_{1}$ score respectively. (2) The ResNet50 backbone with 70$+$pretrain70 achieved the best results, surpassing the original model by 6.40\%, 5.77\%, 7.14\% and 6.46\% in accuracy, precision, recall and F$_{1}$ score respectively. However, while pre-training can improve model accuracy, too much pre-training may cause over-fitting. Besides, due to feature differences between datasets, excessive fine-tuning can also underestimate important features in the pre-trained model, leading to model performance degradation.
\begin{table}[t]\centering
\caption{Performance of ResNet50 model with different training strategies.}
\setlength{\tabcolsep}{0.5mm}{
\begin{tabular}{c|c|c|c|c}
\hline
Method            & Accuracy       & Precision      & Recall         & F$_{1}$ Score       \\ \hline
original          & 78.82\%          & 78.96\%          & 77.55\%  & 78.25\%          \\
40+pretrain\_40   & 81.28\%          & 79.81\%          & 80.61\%  & 80.21\%          \\
40+pretrain\_70  & 81.77\%          & 81.28\%          & 83.67\%     & 82.46\%          \\
70+pretrain\_70 & \textbf{85.22\%} & \textbf{84.73\%} & \textbf{84.69\%} & \textbf{84.71\%} \\
100+pretrain\_70 & 82.76\%          & 82.81\%          & 83.67\%          & 83.24\%          \\
100+pretrain\_100 & 84.24\%          & 83.22\%          & 83.67\%          & 83.44\%          \\ \hline
\end{tabular}}
\label{table5}
\vspace{-6mm}
\end{table}

\section{Limitations and future work}\label{section7}
From the above discussion, we summarise the limitations of this study and possible future work. First, the COVID-19 datasets used in this study are not ideal. Although we used CT and X-ray datasets, which are widely used in COVID-19 detection, the  dataset size is not large due to the challenges of obtaining COVID-labelled data, and there is an imbalance in the number of COVID and normal images of the COVID-Xray-5k dataset. Training and validation on a large dataset collected from the same source will hopefully further enhance the generalisability of the model. We look forward to the emergence of such large datasets, and hope to continue this research at that time. Second, considering the limited datasets, pre-training can be performed on large datasets such as the ChexPert dataset \cite{irvin2019chexpert} in  future, and then transfer learning to our dataset to improve model generalisation and accuracy. Third, clinical studies will be required to validate the effectiveness of this algorithm as an auxiliary tool to help physicians accurately and quickly diagnose COVID-19.

\section{Conclusion}\label{section6}
In this paper, we propose a general method that can be integrated into common classification networks to improve their performance in the detection of COVID-19 from CT scans and X-rays. In detail, our D-SPP module can be used to collect multi-scale image features and context information, and guide subsequent accurate predictions. Our proposed CFD module can maintain the focus of the CNN network on areas of interest related to COVID-19. In addition, the proposed modules can be easily integrated into various deep learning networks to improve their performance. Extensive experiments have been conducted on four COVID-19 CT and X-ray image datasets to evaluate the performance of the proposed method, and the experimental results show the superiority of our approach to other state-of-the-art methods.

\section*{Acknowledgment}

The authors would like to thank the creators of all COVID-19 datasets used in this paper for making the datasets publicly available.

\bibliographystyle{IEEEbib}

\bibliography{cas-refs}

\begin{thebibliography}{10}
\providecommand{\url}[1]{#1}
\csname url@samestyle\endcsname
\providecommand{\newblock}{\relax}
\providecommand{\bibinfo}[2]{#2}
\providecommand{\BIBentrySTDinterwordspacing}{\spaceskip=0pt\relax}
\providecommand{\BIBentryALTinterwordstretchfactor}{4}
\providecommand{\BIBentryALTinterwordspacing}{\spaceskip=\fontdimen2\font plus
\BIBentryALTinterwordstretchfactor\fontdimen3\font minus
  \fontdimen4\font\relax}
\providecommand{\BIBforeignlanguage}[2]{{%
\expandafter\ifx\csname l@#1\endcsname\relax
\typeout{** WARNING: IEEEtran.bst: No hyphenation pattern has been}%
\typeout{** loaded for the language `#1'. Using the pattern for}%
\typeout{** the default language instead.}%
\else
\language=\csname l@#1\endcsname
\fi
#2}}
\providecommand{\BIBdecl}{\relax}
\BIBdecl

\bibitem{su2021vision}
X.~Su, S.~You, J.~Xie, M.~Zheng, F.~Wang, C.~Qian, C.~Zhang, X.~Wang, and
  C.~Xu, ``Vision transformer architecture search,'' \emph{arXiv e-prints}, pp.
  arXiv--2106, 2021.

\bibitem{xu2019automatic}
H.~Xu, X.~Su, Y.~Wang, H.~Cai, K.~Cui, and X.~Chen, ``Automatic bridge crack
  detection using a convolutional neural network,'' \emph{Applied Sciences},
  vol.~9, no.~14, p. 2867, 2019.

\bibitem{woo2018cbam}
S.~Woo, J.~Park, J.-Y. Lee, and I.~S. Kweon, ``Cbam: Convolutional block
  attention module,'' in \emph{Proceedings of the European conference on
  computer vision (ECCV)}, 2018, pp. 3--19.

\bibitem{yang2020imaging}
Q.~Yang, Q.~Liu, H.~Xu, H.~Lu, S.~Liu, and H.~Li, ``Imaging of coronavirus
  disease 2019: a chinese expert consensus statement,'' \emph{European journal
  of radiology}, vol. 127, p. 109008, 2020.

\bibitem{salehi2020coronavirus}
S.~Salehi, A.~Abedi, S.~Balakrishnan, A.~Gholamrezanezhad \emph{et~al.},
  ``Coronavirus disease 2019 (covid-19): a systematic review of imaging
  findings in 919 patients,'' \emph{Ajr Am J Roentgenol}, vol. 215, no.~1, pp.
  87--93, 2020.

\bibitem{huang2020clinical}
C.~Huang, Y.~Wang, X.~Li, L.~Ren, J.~Zhao, Y.~Hu, L.~Zhang, G.~Fan, J.~Xu,
  X.~Gu \emph{et~al.}, ``Clinical features of patients infected with 2019 novel
  coronavirus in wuhan, china,'' \emph{The lancet}, vol. 395, no. 10223, pp.
  497--506, 2020.

\bibitem{ouyang2020dual}
X.~Ouyang, J.~Huo, L.~Xia, F.~Shan, J.~Liu, Z.~Mo, F.~Yan, Z.~Ding, Q.~Yang,
  B.~Song \emph{et~al.}, ``Dual-sampling attention network for diagnosis of
  covid-19 from community acquired pneumonia,'' \emph{IEEE Transactions on
  Medical Imaging}, 2020.

\bibitem{xu2022data}
H.~Xu, X.~Su, S.~You, T.~Huang, F.~Wang, C.~Qian, C.~Zhang, C.~Xu, D.~Wang, and
  A.~Sowmya, ``Data agnostic filter gating for efficient deep networks,'' in
  \emph{ICASSP 2022-2022 IEEE International Conference on Acoustics, Speech and
  Signal Processing (ICASSP)}.\hskip 1em plus 0.5em minus 0.4em\relax IEEE,
  2022, pp. 3503--3507.

\bibitem{soares2020sars}
E.~Soares, P.~Angelov, S.~Biaso, M.~H. Froes, and D.~K. Abe, ``Sars-cov-2
  ct-scan dataset: A large dataset of real patients ct scans for sars-cov-2
  identification,'' \emph{medRxiv}, 2020.

\bibitem{rajpurkar2018deep}
P.~Rajpurkar, J.~Irvin, R.~L. Ball, K.~Zhu, B.~Yang, H.~Mehta, T.~Duan,
  D.~Ding, A.~Bagul, C.~P. Langlotz \emph{et~al.}, ``Deep learning for chest
  radiograph diagnosis: A retrospective comparison of the chexnext algorithm to
  practicing radiologists,'' \emph{PLoS medicine}, vol.~15, no.~11, p.
  e1002686, 2018.

\bibitem{wang2017chestx}
X.~Wang, Y.~Peng, L.~Lu, Z.~Lu, M.~Bagheri, and R.~M. Summers, ``Chestx-ray8:
  Hospital-scale chest x-ray database and benchmarks on weakly-supervised
  classification and localization of common thorax diseases,'' in
  \emph{Proceedings of the IEEE conference on computer vision and pattern
  recognition}, 2017, pp. 2097--2106.

\bibitem{gu2018classification}
X.~Gu, L.~Pan, H.~Liang, and R.~Yang, ``Classification of bacterial and viral
  childhood pneumonia using deep learning in chest radiography,'' in
  \emph{Proceedings of the 3rd International Conference on Multimedia and Image
  Processing}, 2018, pp. 88--93.

\bibitem{wang2020weakly}
X.~Wang, X.~Deng, Q.~Fu, Q.~Zhou, J.~Feng, H.~Ma, W.~Liu, and C.~Zheng, ``A
  weakly-supervised framework for covid-19 classification and lesion
  localization from chest ct,'' \emph{IEEE Transactions on Medical Imaging},
  2020.

\bibitem{he2020sample}
X.~He, X.~Yang, S.~Zhang, J.~Zhao, Y.~Zhang, E.~Xing, and P.~Xie,
  ``Sample-efficient deep learning for covid-19 diagnosis based on ct scans,''
  \emph{medRxiv}, 2020.

\bibitem{wang2020tailored}
L.~Wang, ``A tailored deep convolutional neural network design for detection of
  covid-19 cases from chest x-ray images,'' \emph{arXiv preprint
  arXiv:2003.09871}, 2020.

\bibitem{simonyan2014very}
K.~Simonyan and A.~Zisserman, ``Very deep convolutional networks for
  large-scale image recognition,'' \emph{arXiv preprint arXiv:1409.1556}, 2014.

\bibitem{he2015spatial}
K.~He, X.~Zhang, S.~Ren, and J.~Sun, ``Spatial pyramid pooling in deep
  convolutional networks for visual recognition,'' \emph{IEEE transactions on
  pattern analysis and machine intelligence}, vol.~37, no.~9, pp. 1904--1916,
  2015.

\bibitem{chen2017deeplab}
L.-C. Chen, G.~Papandreou, I.~Kokkinos, K.~Murphy, and A.~L. Yuille, ``Deeplab:
  Semantic image segmentation with deep convolutional nets, atrous convolution,
  and fully connected crfs,'' \emph{IEEE transactions on pattern analysis and
  machine intelligence}, vol.~40, no.~4, pp. 834--848, 2017.

\bibitem{hu2018squeeze}
J.~Hu, L.~Shen, and G.~Sun, ``Squeeze-and-excitation networks,'' in
  \emph{Proceedings of the IEEE conference on computer vision and pattern
  recognition}, 2018, pp. 7132--7141.

\bibitem{su2021k}
X.~Su, S.~You, M.~Zheng, F.~Wang, C.~Qian, C.~Zhang, and C.~Xu, ``K-shot nas:
  Learnable weight-sharing for nas with k-shot supernets,'' in
  \emph{International Conference on Machine Learning}.\hskip 1em plus 0.5em
  minus 0.4em\relax PMLR, 2021, pp. 9880--9890.

\bibitem{su2021bcnet}
X.~Su, S.~You, F.~Wang, C.~Qian, C.~Zhang, and C.~Xu, ``Bcnet: Searching for
  network width with bilaterally coupled network,'' in \emph{Proceedings of the
  IEEE/CVF Conference on Computer Vision and Pattern Recognition}, 2021, pp.
  2175--2184.

\bibitem{su2021prioritized}
X.~Su, T.~Huang, Y.~Li, S.~You, F.~Wang, C.~Qian, C.~Zhang, and C.~Xu,
  ``Prioritized architecture sampling with monto-carlo tree search,'' in
  \emph{Proceedings of the IEEE/CVF Conference on Computer Vision and Pattern
  Recognition}, 2021, pp. 10\,968--10\,977.

\bibitem{su2021locally}
X.~Su, S.~You, T.~Huang, F.~Wang, C.~Qian, C.~Zhang, and C.~Xu, ``Locally free
  weight sharing for network width search,'' \emph{arXiv preprint
  arXiv:2102.05258}, 2021.

\bibitem{bahdanau2014neural}
D.~Bahdanau, K.~Cho, and Y.~Bengio, ``Neural machine translation by jointly
  learning to align and translate,'' \emph{arXiv preprint arXiv:1409.0473},
  2014.

\bibitem{xu2015show}
K.~Xu, J.~Ba, R.~Kiros, K.~Cho, A.~Courville, R.~Salakhudinov, R.~Zemel, and
  Y.~Bengio, ``Show, attend and tell: Neural image caption generation with
  visual attention,'' in \emph{International conference on machine learning},
  2015, pp. 2048--2057.

\bibitem{rush2015neural}
A.~M. Rush, S.~Chopra, and J.~Weston, ``A neural attention model for
  abstractive sentence summarization,'' \emph{arXiv preprint arXiv:1509.00685},
  2015.

\bibitem{zhao2017pyramid}
H.~Zhao, J.~Shi, X.~Qi, X.~Wang, and J.~Jia, ``Pyramid scene parsing network,''
  in \emph{Proceedings of the IEEE conference on computer vision and pattern
  recognition}, 2017, pp. 2881--2890.

\bibitem{zhou2014object}
B.~Zhou, A.~Khosla, A.~Lapedriza, A.~Oliva, and A.~Torralba, ``Object detectors
  emerge in deep scene cnns,'' \emph{arXiv preprint arXiv:1412.6856}, 2014.

\bibitem{zoph2018learning}
B.~Zoph, V.~Vasudevan, J.~Shlens, and Q.~V. Le, ``Learning transferable
  architectures for scalable image recognition,'' in \emph{Proceedings of the
  IEEE conference on computer vision and pattern recognition}, 2018, pp.
  8697--8710.

\bibitem{zagoruyko2016wide}
S.~Zagoruyko and N.~Komodakis, ``Wide residual networks,'' \emph{arXiv preprint
  arXiv:1605.07146}, 2016.

\bibitem{selvaraju2017grad}
R.~R. Selvaraju, M.~Cogswell, A.~Das, R.~Vedantam, D.~Parikh, and D.~Batra,
  ``Grad-cam: Visual explanations from deep networks via gradient-based
  localization,'' in \emph{Proceedings of the IEEE international conference on
  computer vision}, 2017, pp. 618--626.

\bibitem{minaee2020deep}
S.~Minaee, R.~Kafieh, M.~Sonka, S.~Yazdani, and G.~J. Soufi, ``Deep-covid:
  Predicting covid-19 from chest x-ray images using deep transfer learning,''
  \emph{arXiv preprint arXiv:2004.09363}, 2020.

\bibitem{irvin2019chexpert}
J.~Irvin, P.~Rajpurkar, M.~Ko, Y.~Yu, S.~Ciurea-Ilcus, C.~Chute, H.~Marklund,
  B.~Haghgoo, R.~Ball, K.~Shpanskaya \emph{et~al.}, ``Chexpert: A large chest
  radiograph dataset with uncertainty labels and expert comparison,'' in
  \emph{Proceedings of the AAAI Conference on Artificial Intelligence},
  vol.~33, 2019, pp. 590--597.

\bibitem{alshazly2020explainable}
H.~Alshazly, C.~Linse, E.~Barth, and T.~Martinetz, ``Explainable covid-19
  detection using chest ct scans and deep learning,'' \emph{arXiv preprint
  arXiv:2011.05317}, 2020.

\bibitem{jaiswal2020classification}
A.~Jaiswal, N.~Gianchandani, D.~Singh, V.~Kumar, and M.~Kaur, ``Classification
  of the covid-19 infected patients using densenet201 based deep transfer
  learning,'' \emph{Journal of Biomolecular Structure and Dynamics}, pp. 1--8,
  2020.

\bibitem{panwar2020deep}
H.~Panwar, P.~Gupta, M.~K. Siddiqui, R.~Morales-Menendez, P.~Bhardwaj, and
  V.~Singh, ``A deep learning and grad-cam based color visualization approach
  for fast detection of covid-19 cases using chest x-ray and ct-scan images,''
  \emph{Chaos, Solitons \& Fractals}, vol. 140, p. 110190, 2020.

\bibitem{yazdani2020covid}
S.~Yazdani, S.~Minaee, R.~Kafieh, N.~Saeedizadeh, and M.~Sonka, ``Covid ct-net:
  Predicting covid-19 from chest ct images using attentional convolutional
  network,'' \emph{arXiv preprint arXiv:2009.05096}, 2020.

\bibitem{wang2020contrastive}
Z.~Wang, Q.~Liu, and Q.~Dou, ``Contrastive cross-site learning with redesigned
  net for covid-19 ct classification,'' \emph{IEEE Journal of Biomedical and
  Health Informatics}, vol.~24, no.~10, pp. 2806--2813, 2020.

\bibitem{ragab2020fusi}
D.~A. Ragab and O.~Attallah, ``Fusi-cad: Coronavirus (covid-19) diagnosis based
  on the fusion of cnns and handcrafted features,'' \emph{PeerJ Computer
  Science}, vol.~6, p. e306, 2020.

\bibitem{pathak2020deep}
Y.~Pathak, P.~K. Shukla, and K.~Arya, ``Deep bidirectional classification model
  for covid-19 disease infected patients,'' \emph{IEEE/ACM Transactions on
  Computational Biology and Bioinformatics}, 2020.

\bibitem{saqib2020covid19}
M.~Saqib, S.~Anwar, A.~Anwar, M.~Blumenstein \emph{et~al.}, ``Covid19 detection
  from radiographs: Is deep learning able to handle the crisis?'' 2020.

\bibitem{ewen2020targeted}
N.~Ewen and N.~Khan, ``Targeted self supervision for classification on a small
  covid-19 ct scan dataset,'' \emph{arXiv preprint arXiv:2011.10188}, 2020.

\bibitem{babukarthik2020prediction}
R.~Babukarthik, V.~A.~K. Adiga, G.~Sambasivam, D.~Chandramohan, and
  J.~Amudhavel, ``Prediction of covid-19 using genetic deep learning
  convolutional neural network (gdcnn),'' \emph{IEEE Access}, vol.~8, pp.
  177\,647--177\,666, 2020.

\bibitem{ozdemir2020weighted}
{\"O}.~{\"O}zdemir and E.~B. S{\"o}nmez, ``Weighted cross-entropy for
  unbalanced data with application on covid x-ray images,'' in \emph{2020
  Innovations in Intelligent Systems and Applications Conference (ASYU)}.\hskip
  1em plus 0.5em minus 0.4em\relax IEEE, 2020, pp. 1--6.

\end{thebibliography}

\end{document}